\documentclass[]{article}  
\usepackage{graphicx}
\usepackage{amssymb}
\usepackage{amsmath}
\usepackage{times}
\usepackage{color}

\usepackage{url}
\usepackage[numbers,sort&compress]{natbib}
\usepackage{bbm}
\usepackage[table]{xcolor}
\usepackage{tikz}

\definecolor{MyBlue}{cmyk}{0.2,0 ,0 ,0}

\begin{document} 
\title{Counterintuitive properties of the fixation time in network-structured populations} 
\author{Laura Hindersin and Arne Traulsen \\
Department of Evolutionary Theory \\
Max Planck Institute for Evolutionary Biology \\
Pl\"on, Germany}
\date{\today}
\maketitle

\begin{abstract}
Evolutionary dynamics on graphs can lead to many interesting and counterintuitive findings. 
We study the Moran process, a discrete time birth-death process, that describes the invasion of a mutant type into a population of wild-type individuals.
Remarkably, the fixation probability of a single mutant is the same on all regular networks.
But non-regular networks can increase or decrease the fixation probability.
While the time until fixation formally depends on the same transition probabilities as the fixation probabilities, there is
no obvious relation between them. 
For example, an amplifier of selection, which increases the fixation probability and thus decreases the number of mutations needed until one of them is successful, can at the same
time slow down the process of fixation. 
Based on small networks, we show analytically that 
(i) the time to fixation can decrease when links are removed from the network and
(ii) the node providing the best starting conditions in terms of the shortest fixation time depends on the fitness of the mutant.
Our results are obtained analytically on small networks, but numerical simulations show that
they are qualitatively valid even in much larger populations. 
\end{abstract}

\section{Introduction}

Most analytical results for evolutionary dynamics have been obtained for well-mixed populations,
regular networks or deme-structures. 
However, the consideration of non-regular networks has shown that there is a wealth of evolutionary phenomena that is not captured by these approaches. 
For example, while the fixation probability of a single mutant is the same on all regular networks \citep{lieberman:Nature:2005}, 
some non-regular networks can increase this probability and serve as amplifiers of selection, 
or decrease it and serve as suppressors of selection.
It seems to be tempting to use amplifiers of selection to speed up experimental evolution, as the probability of the fixation of beneficial mutants could be increased. 
However, it turns out that amplifiers of selection can at the same time slow down the time until fixation.
Thus, such an approach would have the drawback that while the time until a mutant appears is decreased, the time until it takes over is increased \citep{broom:PRSA:2010,frean:PRSB:2013}. 

Evolutionary dynamics on network structures can also serve as a powerful abstraction when studying the somatic evolution of cancer \citep{nowak:PNAS:2003,michor:PRSB:2003,komarova:MB:2006}. In this context, the idea is that directed networks can substantially decrease the number of cells at risk and thus inhibit the accumulation of cancer mutations.

While most previous work on network structured populations has focused on the fixation probability 
\citep{nowak:PNAS:2003,lieberman:Nature:2005,traulsen:JTB:2005,antal:PRL:2006,broom:JIM:2009,broom:JSTP:2011,monk:PRSA:2014}, the time to fixation has received considerably less attention so far \citep{broom:PRSA:2010,broom:book:2013,frean:PRSB:2013}. 
Several questions that appear somewhat obvious are still open:
does the time to fixation always increase when a link is removed from a network?
Do amplifiers of selection always change the time to fixation?

We study constant selection, where the fitness does not depend on the frequencies of the types. 
There exists a huge body of closely related research on evolutionary game theory on graphs, e.g. \citep{nowak:Nature:1992b,abramson:PRE:2001,hauert:IJBC:2002,santos:PRL:2005,ohtsuki:Nature:2006,szabo:PR:2007,roca:PLR:2009,perc:BioSys:2010,nowak:PTRSB:2010,broom:book:2013}. 
It has been shown there that the graph structure can substantially affect evolutionary game dynamics. 
But as the fixation time on networks already leads to interesting and counterintuitive results for constant selection, we focus on this case.

As the number of possible states in non-regular networks rapidly increases with population size $N$, we focus on the smallest population size that allows us to obtain any interesting results, $N=4$.
We consider all six possible undirected networks in detail and calculate 
(i) the probability of fixation
(ii) the time to fixation
(iii) the sojourn times
depending on the fitness of the mutant. 
This approach illustrates that the time to fixation can actually increase when a link is added to the network.
It particularly allows to infer in which states the system spends this additional time. 
Moreover, our approach shows that the optimal starting point for a novel mutation in terms of its fixation time also depends on the fitness of this mutation.

\subsection{The Moran process in well-mixed populations}
 
The Moran process is a birth-death process in a well-mixed population \citep{moran:book:1962}.
We start from $N-1$ wild-type individuals with fitness $1$ and one mutant with fitness $r>0$. 
If $r>1$, the mutant is advantageous compared to the wild-type, whereas $r<1$ implies a disadvantage. One can also study neutral evolution, where $r=1$. 
At each time step, one of the $N$ individuals is selected for birth with probability proportional to its fitness. 
This individual gives birth to an identical offspring which replaces another randomly chosen individual. 

The probabilities of increasing and decreasing the number of mutants are given by a tridiagonal transition matrix $\mathbf{T}_{(N+1) \times (N+1)}$.
The probability to increase the number of mutants by one is given by
\begin{equation}
 T_{i,i+1} = \frac{ri}{ri+N-i} \cdot \frac{N-i}{N-1}\ , \label{eq:tPlus} 
\end{equation}
for $0 \leq i \leq N$. 
The probability to decrease the number of mutants from $i$ to $i-1$ in one time step is
\begin{equation}
 T_{i,i-1} = \frac{N-i}{ri+N-i} \cdot \frac{i}{N-1}\ , \label{eq:tMinus}
\end{equation}
for $0\leq i \leq N$. 

We assume a mutation rate of zero, which is reflected by $T_{0,1} =T_{N,N-1}= 0$. 
As $T_{0,-1}=T_{N,N+1} =  0$, the boundaries $i=0$ and $i=N$ are absorbing. 

If a mutant replaces another mutant or a wild-type replaces another wild-type, the population does not change. 
This happens with probability $T_{i,i} = 1- T_{i,i+1} - T_{i,i-1}$.

As the ratio of the transition probabilities is 
\begin{equation}
\frac{T_{i,i-1}}{T_{i,i+1}}=\frac{1}{r}
\end{equation}
for $1\leq i \leq N-1$, the fixation probability $\Phi_i^N$ for $i$ mutants in a well-mixed population is given by \citep{karlin:book:1975,nowak:book:2006,traulsen:bookchapter:2009}
\begin{equation}
\Phi_i^N = 
\frac{1+\sum\limits_{k=1}^{i-1} \prod\limits_{l=1}^{k} \frac{T_{l,l-1}}{T_{l,l+1}}}{1+\sum\limits_{k=1}^{N-1} \prod\limits_{l=1}^{k} \frac{T_{l,l-1}}{T_{l,l+1}}}
=
\frac{1-\frac{1}{r^i}}{1-\frac{1}{r^N}} \ .
\end{equation}

The time which one mutant needs to take over the whole population, given that it does succeed, is called conditional fixation time because it is conditioned on the success of the mutants. 
By contrast, the unconditional fixation time measures how long it takes for the mutants to either go extinct or fixate, starting from one mutant.
For birth-death processes (and the Moran process in particular), the expected conditional fixation time $\tau_1^N$ is given by
\citep{karlin:book:1975,nowak:book:2006,traulsen:bookchapter:2009}
\begin{equation}
 \tau_1^N = \sum_{k=1}^{N-1}{\sum_{l=1}^k{\frac{\Phi_l^N}{T_{l,l+1}}\prod_{m=l+1}^k{\frac{T_{m,m-1}}{T_{m,m+1}}}}} \quad .
 \label{eq:condTime}
\end{equation}

\subsection{The Moran process in structured populations}

Population structure can be introduced into the Moran model by assuming that individuals are represented by the nodes in a network and assuming that reproduction and replacement is only possible between connected nodes, see 
\citep[Chapter 8]{lieberman:Nature:2005,nowak:book:2006}. 
The fixation probability can then be assessed in the following way: 
individuals of the wild-type with fitness $1$ are placed on the nodes of a network. 
A single mutant with fitness $r$ is placed on one of the nodes at random. 
In each time step, one individual is chosen for birth with probability proportional to its fitness. 
Then one of its neighbours is chosen at random to be replaced by the new offspring. 
Thus, the links of a node determine into which of the neighbouring sites the individual on that node can reproduce.
Throughout our work, we only consider connected undirected networks with equal weights. 
The standard Moran process corresponds to the complete network, where every node is adjacent to all other nodes, which implies that the probability of being replaced is equal for all individuals.

On isothermal networks, where every node has the same number of neighbours in the case of undirected networks, it has been shown in \citep{lieberman:Nature:2005} that the fixation probability is the same as in the well-mixed population. 
Examples for isothermal networks are the ring and the two-dimensional lattice with periodic boundary conditions.
But in general, the probability of replacement can vary crucially between different nodes,
resulting in fixation probabilities that are distinct from the well-mixed population. 
Non-isothermal networks that increase (decrease) the fixation probability for advantageous mutants and decrease (increase) it for disadvantageous mutants are called amplifiers (suppressors) of selection. 

\subsection{A general approach to calculate probabilities and times of fixation}

\label{subsec:Grinstead}
While the analytical results for birth-death processes can be tailored to some network-structures \cite{broom:PRSA:2008,frean:PRSB:2013}, this does not always work. 
For assessing absorption probabilities and times of Markov chains, a more general approach, which is typically used in numerical considerations, described in \citep[Chapter 11]{grinstead:book:1997} can be exploited \citep[see also][for an application in another context]{hauser:JTB:2014}.

Let $\textbf{T}_{s\times s}$ be the transition matrix of a discrete-time Markov chain with at least one absorbing state. 
We renumber the states such that the $t$ transient states are first and the $a$ absorbing states are last, where $t+a=s$ is the total number of states (for our problem, we will always consider the case of two types, which implies $a=2$). 
The transition matrix now has the following canonical form:
\begin{equation}
\mathbf{T}_{s\times s}\ = \ \left(
 \begin{array}{cc}
  \mathbf{Q}& \mathbf{R}\\
 \mathbf{0}& \mathbf{I}\\
   \end{array} \right)
   \ ,
\end{equation}
where $\mathbf{Q}_{t\times t}$ consists of the transition probabilities between transient states and $\mathbf{R}_{t\times a}$ describes the transitions from the transient into the absorbing states. 
As transitions are not possible from an absorbing state to a transient state, the lower left block is zero. 
Once absorbed, the process will stay there forever, therefore the lower right block of $\mathbf{T}$ is an identity matrix $\mathbf{I}_{a\times a}$.
For a starting distribution $x_{1\times s}$, the product $x\mathbf{T}^m$ gives the distribution after exactly $m$ time steps.
For $m \to \infty$, we can recover the fixation probabilities from this. 

Let us call $\mathbf{F} = \sum_{n=0}^{\infty}{\mathbf{Q}^n} = (\mathbf{I}-\mathbf{Q})^{-1} $ the fundamental matrix of the Markov chain. 
The $i,j$-th entry of $\mathbf{F}$, given by
\begin{equation}
F_{i,j} = \left( \left( \mathbf{I}-\mathbf{Q} \right)^{-1} \right)_{i,j}
\end{equation} 
is the expected sojourn time in the transient state $j$, given that the process started in the transient state $i$ \citep[Chapter 11]{grinstead:book:1997}.

Multiplying $\mathbf{F}$ with the transition probabilities to the absorbing states provides the absorption probabilities. 
The absorption probability in state $j$ after starting in state $i$,
$\phi_{i,j}$, is the $ij$-th entry of $\Phi= \mathbf{F} \cdot \mathbf{R}$, 
\begin{equation}
\phi_{i,j} = \left( \left( \mathbf{I}-\mathbf{Q} \right)^{-1} \mathbf{R} \right)_{i,j}.
\end{equation}

To assess the total time the process spends before absorption, we look at the so-called unconditional fixation time.
The unconditional fixation time, which is the expected number of time steps before the process is absorbed in either of the absorbing states can be calculated from the expected sojourn time.
Summing over the $i$-th row of $\mathbf{F}$ yields the unconditional fixation time, after starting in state $i$. 
For $N-1$ transient states, this is given by
\begin{equation}
\tau_i = \sum_{j=1}^{N-1} F_{i,j} \ .
\end{equation}

To calculate the conditional fixation time from the unconditional sojourn times we use an approach described in \citep{ewens:TPB:1973,altrock:JTB:2012}.
If $\Phi_i^N$ is the fixation probability to reach state $N$ starting in state $i$ 
and $F_{i,j}$ is the unconditional sojourn time in state $j$ after starting in state $i$, 
then the conditional fixation time of the process on any network, starting in state $i$, is given by
\begin{equation}
\tau_i^N = \sum_{j=1}^{N-1}{\left( \frac{\Phi_j^N}{\Phi_i^N} \cdot F_{i,j}\right) }\ .
\label{condfixtime}
\end{equation}
This allows us to go beyond unconditional times even on heterogeneous networks.

\section{Results}

\subsection{Analytical results for small networks}
\begin{figure}[h!]
\centering
\includegraphics[width=0.7\textwidth]{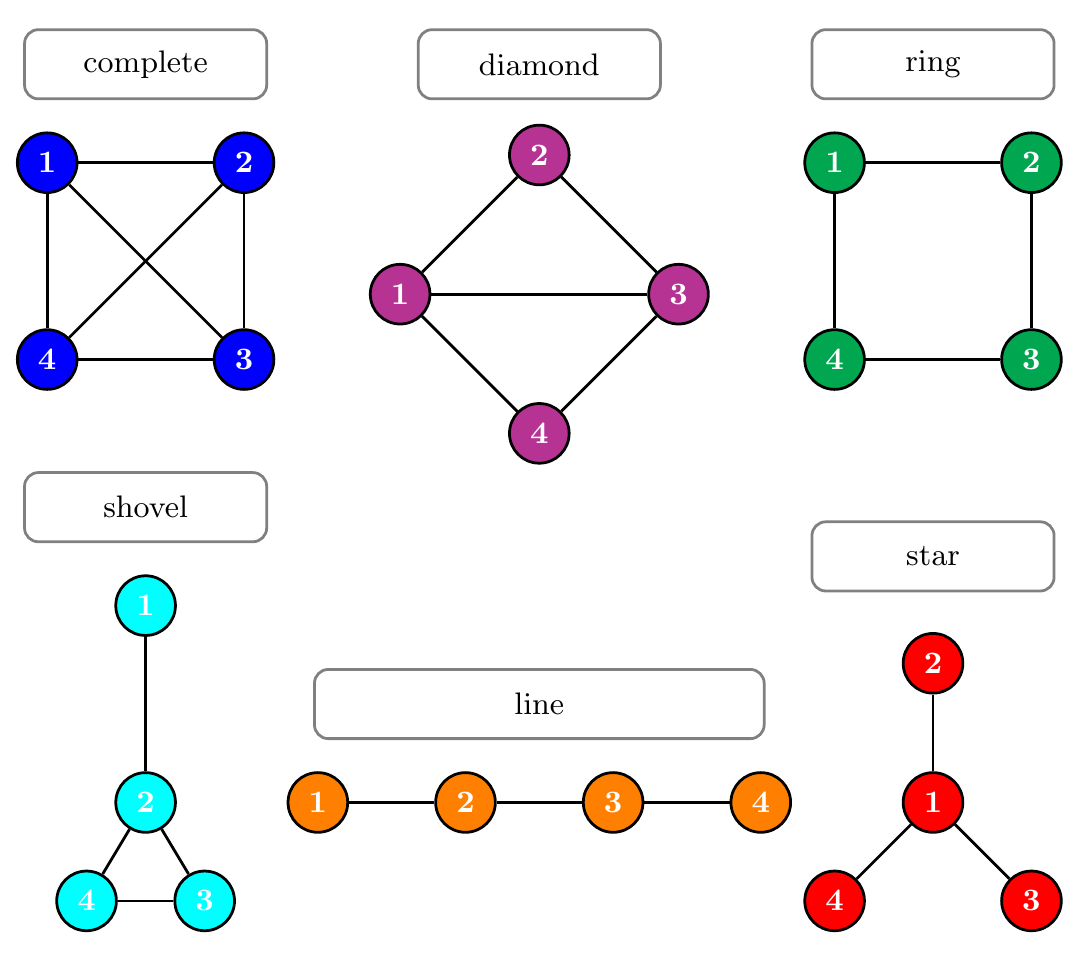}
\caption{The six possible connected undirected networks of size $N=4$. 
The complete graph (which corresponds to the well-mixed population) and the ring are isothermal (homogeneous) networks, implying that 
the fixation probabilities on these networks are identical.
The other four networks are degree heterogeneous and thus not isothermal.
}
\label{fig:AllGraphsSize4}
\end{figure}

First, we analyse small networks that allow a fully analytical approach. 
There are six different connected undirected graphs with four nodes, illustrated in Fig.\ \ref{fig:AllGraphsSize4}.

Out of these six networks in Fig.\ \ref{fig:AllGraphsSize4}, only the complete graph and the ring are isothermal. 
On the other four networks, the nodes vary in degree, i.e. number of neighbours.
The diamond has two nodes with degree two and two nodes with degree three.
On the shovel, there are three types of nodes: having either one, two or three neighbours. 
The least possible number of links in a connected network of size four is three links. The two associated networks are called the line and the star.
The two outer nodes of the line have one neighbour and the two inner nodes have two neighbours.
On the star, the nodes vary even more in degree: the centre node has three neighbours whereas the three leaf nodes are only connected to the centre.

\subsubsection{States and transition matrices}
To calculate the fixation probability, we first look at the different possible states and the transitions between states. 
Then we rearrange the states in the transition matrix as discussed in Section 1\ref{subsec:Grinstead}, such that the transient states are first. \\
  
\begin{figure}[h!]
 \centering
 \includegraphics[width=\textwidth]{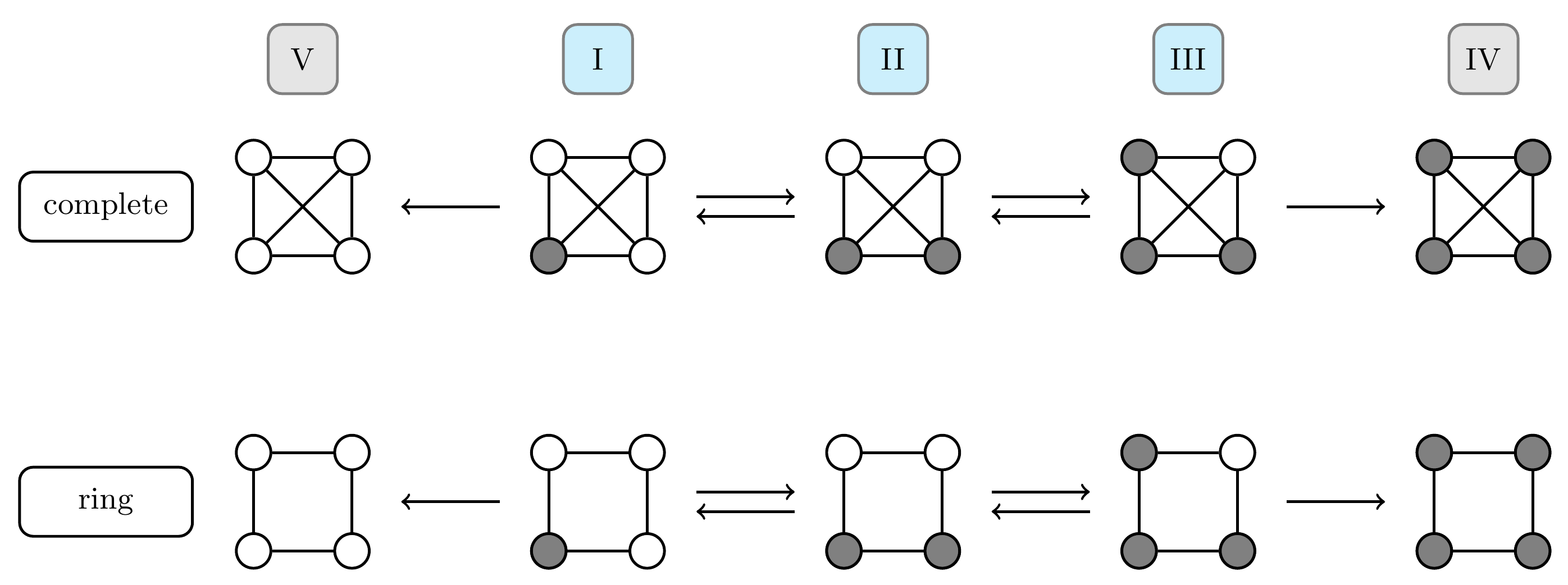}
 \caption{The five states of a Markov chain on a complete graph and a ring of size four. 
 Gray nodes indicate mutants, whereas white nodes represent wild-type individuals. 
 The arrows show possible transitions between states of the chain in one time step. 
 The process can also stay in the same state whenever a mutant or wild-type individual replaces one of its own kind.
 The process starts at state I and moves around on the state space until it reaches one of the absorbing states IV or V.
 Note that for these two networks, the same transitions are possible, but the transition probabilities are different.}
 \label{fig:statesMixed}
\end{figure}

\noindent \textbf{Complete graph and ring of size four} \\
Let I, II, III and IV be the states with $1,2,3$ and $4$ mutants, respectively, and V the state with only wild-type individuals. 
The states of this Markov chain are shown in Fig.\ \ref{fig:statesMixed}. 
Transient state numbers are highlighted in blue, whereas absorbing states are shaded in gray.

In Fig.\ \ref{fig:statesMixed}, the five different states of the Moran process on a well-mixed population and a ring of size four are displayed. State I is the initial state. If the absorbing state IV is reached, this means that the mutant reached fixation in the population.
Due to the special structure of the ring, mutants can only invade in a cluster. 
Therefore a change in the number of mutants can only happen if nodes at the boundary of the mutant cluster are selected for birth.

The canonical form of the transition matrix for the process on the complete graph is given in Eq.\ \eqref{eq:transitionMatrixMix}. Transient states are highlighted in blue and absorbing states in light gray.

 \begin{equation}
 \mathbf{T}_{\textrm{mix}}= \quad
 \begin{array}{l|ccc|cc}
 \textrm{state}& \text{I} & \text{II} & \text{III} & \text{IV} & \text{V} \\ 
 \hline
 \text{I}&   \cellcolor{MyBlue}
 \frac{2}{r + 3} &  \cellcolor{MyBlue}\frac{r}{r + 3} &  \cellcolor{MyBlue}0 & \cellcolor{gray!20}0 & \cellcolor{gray!20}\frac{1}{r + 3} \\
 \text{II}&  \cellcolor{MyBlue}\frac{2}{3(r + 1)} &  \cellcolor{MyBlue}\frac{1}{3} &  \cellcolor{MyBlue}\frac{2 r}{3(r + 1)} & \cellcolor{gray!20}0 & \cellcolor{gray!20}0 \\
 \text{III}&  \cellcolor{MyBlue}0 &  \cellcolor{MyBlue}\frac{1}{3 r + 1} &  \cellcolor{MyBlue}\frac{2 r}{3 r + 1} & \cellcolor{gray!20}\frac{r}{3 r + 1} & \cellcolor{gray!20}0\\
 \hline 
 \text{IV}& 0 & 0 & 0 & 1 & 0 \\
 \text{V}& 0 & 0 & 0 & 0 & 1
 \end{array}
 \label{eq:transitionMatrixMix}
 \end{equation}
The diagonal of the transition matrix $\mathbf{T}_{\textrm{mix}}$ is positive. 
The Moran process stays in the same state, meaning that the number of mutants does not change whenever a mutant replaces a mutant or a wild-type individual replaces another wild-type individual.

With the approach given in Section 1\ref{subsec:Grinstead}, we reproduce the well-known fixation probability of a mutant in the well-mixed population \citep{karlin:book:1975,nowak:book:2006}:
\begin{equation}
\Phi_{1\  \textrm{mix}}^N \ =\  \frac{1- \frac{1}{r}}{1- \frac{1}{r^4}} \ .
\label{eq:mixFixProb}
\end{equation}
For the ring, we obtain the canonical transition matrix
 \begin{equation}
 \mathbf{T}_{\textrm{ring}}=\quad\begin{array}{l|ccc|cc}
 \textrm{state} & \text{I} & \text{II} & \text{III} & \text{IV} & \text{V} \\
 \hline
 \text{I}& \cellcolor{MyBlue}\frac{2}{r + 3} & \cellcolor{MyBlue}\frac{r}{r + 3} & \cellcolor{MyBlue}0 & \cellcolor{gray!20}0 & \cellcolor{gray!20}\frac{1}{r + 3} \\
 \text{II}& \cellcolor{MyBlue}\frac{1}{2(r + 1)} & \cellcolor{MyBlue}\frac{1}{2} & \cellcolor{MyBlue}\frac{r}{2(r + 1)} & \cellcolor{gray!20}0 & \cellcolor{gray!20}0 \\
 \text{III}& \cellcolor{MyBlue}0 & \cellcolor{MyBlue}\frac{1}{3 r + 1} & \cellcolor{MyBlue}\frac{2 r}{3 r + 1} & \cellcolor{gray!20}\frac{r}{3 r + 1} & \cellcolor{gray!20}0\\
 \hline
 \text{IV}& 0 & 0 & 0 & 1 & 0\\
 \text{V}& 0 & 0 & 0 & 0 & 1
 \end{array}
 \label{eq:transitionMatrixRing}
 \end{equation}

Recall that the ring is isothermal. 
Hence, the fixation probability must be the same as for the well-mixed population and given by Eq.\ \eqref{eq:mixFixProb} as well. 
Indeed, the ratio $\frac{T_{i,i-1}}{T_{i,i+1}}=\frac{1}{r}$ remains unchanged for transient states $1\leq i \leq N-1$. 

However, the second line of $\mathbf{T}_{\textrm{ring}}$ in Eq.\ \eqref{eq:transitionMatrixRing} is different from the second line of $\mathbf{T}_{\textrm{mix}}$ in Eq.\ \eqref{eq:transitionMatrixMix}. 
For example, whenever there are two mutants on the ring network, the probability to stay with two mutants in the next time step is $T_{\text{II},\text{II}}=\frac{1}{2}$, whereas the corresponding probability to stay at the state with two mutants on the complete graph, is only $T_{\text{II},\text{II}}=\frac{1}{3}$.
This indicates that fixation takes longer on the ring, because the probability of no change in state II is higher, see \citep{frean:PRSB:2013}. \\

\begin{figure}[h!]
 \centering
  \includegraphics[width=0.9\textwidth]{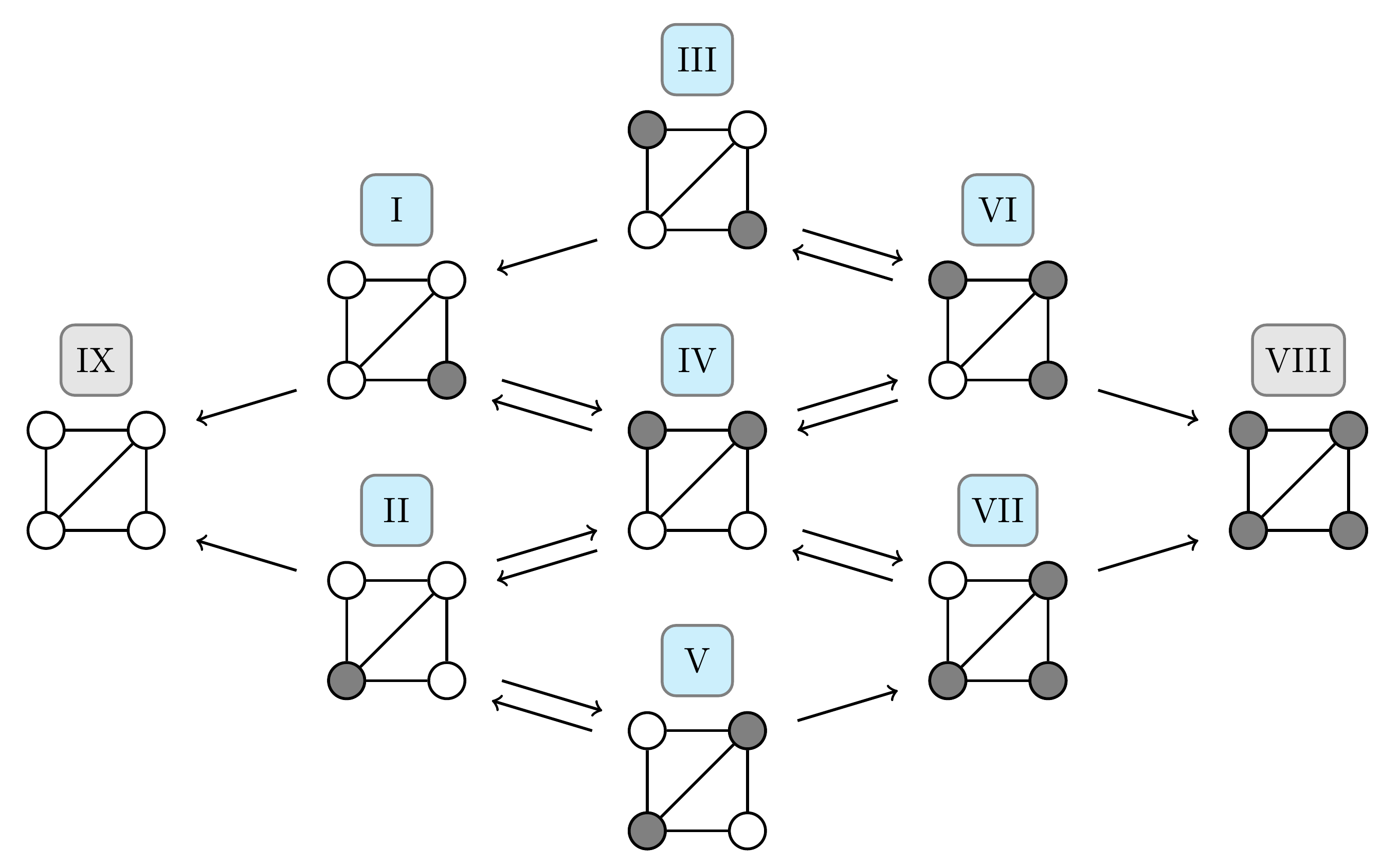}
  \caption{Owing to the different degrees of the nodes, the diamond has more possible states than the complete graph. 
  There are two nodes with degree three and two nodes with degree two. 
  Therefore, one must distinguish between those two types of nodes. 
  For example, this leads to the distinction between states I and II, which would be the same, if all nodes had the same degree.
  Note that even between the transient states I,II,$\dots$,VII not all transitions are bidirected. 
  The transition probabilities $T_{\text{I},\text{III}}$ and $T_{\text{VII},\text{V}}$ are zero and of course, vertical transitions are not possible.}
    \label{graph:StatesFiveLinks}
 \end{figure}

\noindent \textbf{Diamond} \\
Next, let us consider the diamond, which has 9 states I,II,$\dots$,IX shown in Fig.\ \ref{graph:StatesFiveLinks}. 
There are several possible mutant-node configurations in this network, such that the states are no longer just determined by the number of mutants. 
Instead, they are also determined by the degree of the respective node. 
 
Owing to the larger state space, the process on the diamond is not a simple birth-death process. Thus, the transition matrix does not have a tridiagonal shape and the conditional fixation time is not given by Eq.\ \eqref{eq:condTime}, but has to be computed from Eq.\ \eqref{condfixtime}.
The canonical form of the transition matrix for the diamond is given by

\small
 \begin{equation*}
 \begin{array}{l | ccccccc | cc}
  \textrm{state} & \text{I} & \text{II} & \text{III} & \text{IV} & \text{V} & \text{VI} & \text{VII} & \text{VIII} & \text{IX}\\
  \hline
  \text{I}& \cellcolor{MyBlue}\frac{7}{3 (r + 3)} & \cellcolor{MyBlue}0 & \cellcolor{MyBlue}0 & \cellcolor{MyBlue}\frac{r}{r + 3} & \cellcolor{MyBlue}0 & \cellcolor{MyBlue}\cellcolor{MyBlue}0 & \cellcolor{MyBlue}0 & \cellcolor{gray!20}0 & \cellcolor{gray!20}\frac{2}{3(r + 3)} \\
  \text{II} & \cellcolor{MyBlue}0 & \cellcolor{MyBlue}\frac{5}{3(r + 3)} & \cellcolor{MyBlue}0 & \cellcolor{MyBlue}\frac{2r}{3(r + 3)} & \cellcolor{MyBlue}\frac{r}{3(r + 3)} & \cellcolor{MyBlue}0 & \cellcolor{MyBlue}0 & \cellcolor{gray!20}0 & \cellcolor{gray!20}\frac{4}{3(r + 3)} \\
  \text{III} & \cellcolor{MyBlue}\frac{2}{3(r + 1)} & \cellcolor{MyBlue}0 & \cellcolor{MyBlue}\frac{1}{3(r + 1)} & \cellcolor{MyBlue}0 & \cellcolor{MyBlue}0 & \cellcolor{MyBlue}\frac{r}{r + 1} & \cellcolor{MyBlue}0 & \cellcolor{gray!20}0 & \cellcolor{gray!20}0 \\
  \text{IV} & \cellcolor{MyBlue}\frac{5}{12(r + 1)} & \cellcolor{MyBlue}\frac{1}{6(r + 1)} & \cellcolor{MyBlue}0 & \cellcolor{MyBlue}\frac{5 }{12} & \cellcolor{MyBlue}0 & \cellcolor{MyBlue}\frac{r}{6(r + 1)} & \cellcolor{MyBlue}\frac{5r}{12(r + 1)} & \cellcolor{gray!20}0 & \cellcolor{gray!20}0 \\
  \text{V} & \cellcolor{MyBlue}0 & \cellcolor{MyBlue}\frac{1}{r + 1} & \cellcolor{MyBlue}0 & \cellcolor{MyBlue}0 & \cellcolor{MyBlue}\frac{r}{3 (r + 1)} & \cellcolor{MyBlue}0 & \cellcolor{MyBlue}\frac{2r}{3(r + 1)} & \cellcolor{gray!20}0 & \cellcolor{gray!20}0 \\
  \text{VI} & \cellcolor{MyBlue}0 & \cellcolor{MyBlue}0 & \cellcolor{MyBlue}\frac{1}{3(3 r + 1)} & \cellcolor{MyBlue}\frac{2}{3(3 r + 1)} & \cellcolor{MyBlue}0 & \cellcolor{MyBlue}\frac{5r}{3(3 r + 1)} & \cellcolor{MyBlue}0 & \cellcolor{gray!20}\frac{4r}{3(3 r + 1)} & \cellcolor{gray!20}0 \\
  \text{VII} & \cellcolor{MyBlue}0 & \cellcolor{MyBlue}0 & \cellcolor{MyBlue}0 & \cellcolor{MyBlue}\frac{1}{3r + 1} & \cellcolor{MyBlue}0 & \cellcolor{MyBlue}0 & \cellcolor{MyBlue}\frac{7r}{3(3 r + 1)} & \cellcolor{gray!20}\frac{2r}{3(3 r + 1)} & \cellcolor{gray!20}0 \\
  \hline
  \text{VIII} & 0 & 0 & 0 & 0 & 0 & 0 & 0 & 1 & 0 \\
  \text{IX} & 0 & 0 & 0 & 0 & 0 & 0 & 0 & 0 & 1
  \end{array}
 \end{equation*}
\normalsize
  
The state space and transition matrix of the other three networks of size four (shovel, line and star, see Fig.\ \ref{fig:AllGraphsSize4}) can be analysed in a similar fashion, but a numerical approach is often more efficient to implement. 
In the following section, we use the transition matrices to compute the fixation probabilities on these networks.

\subsubsection{Fixation probabilities}

\begin{figure}[h!]
  \centering
   \includegraphics[width=0.8\textwidth]{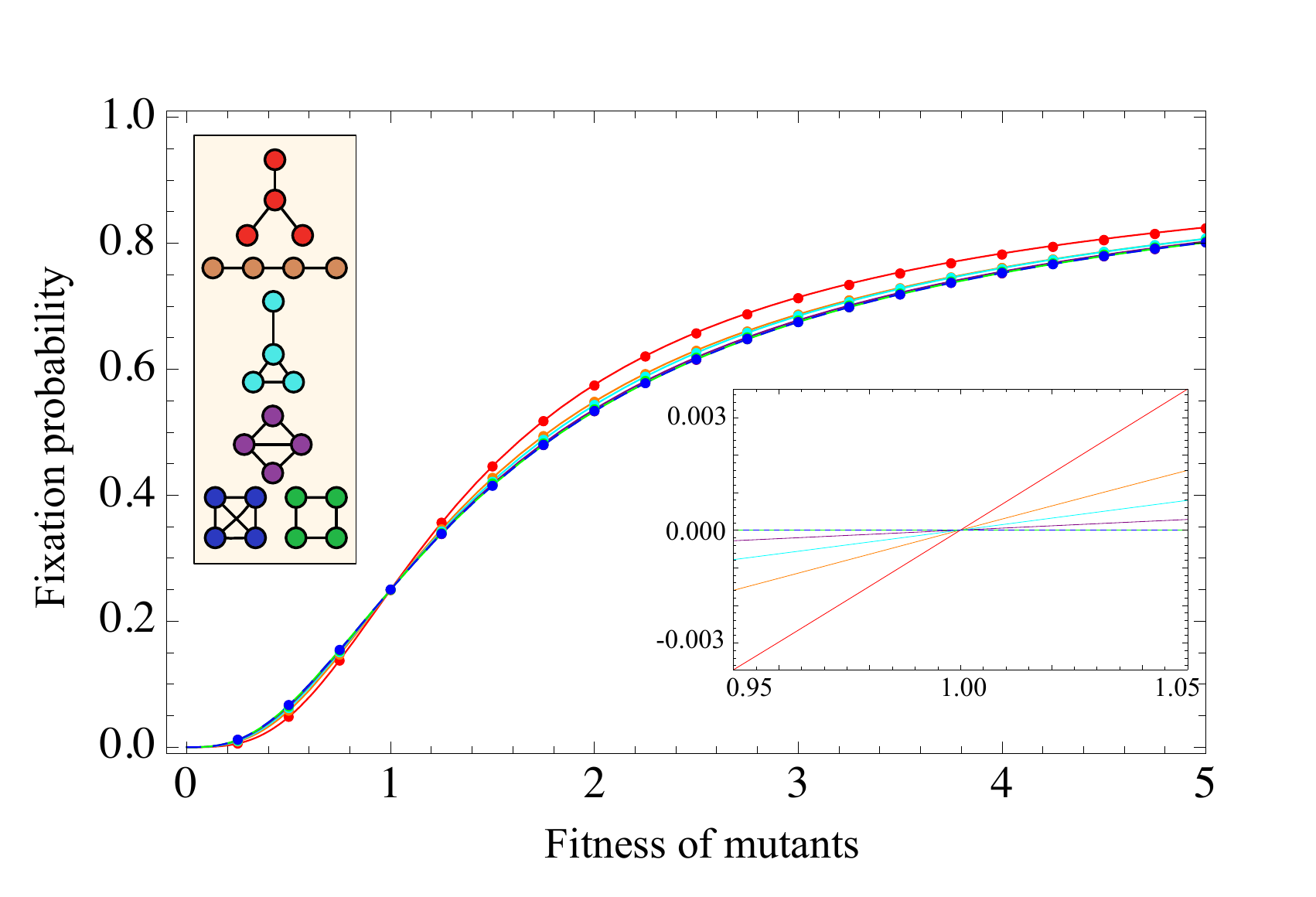} 
 \caption{
Fixation probability for the six structures of size four. 
Lines show analytical results, whereas every dot represents the frequency of fixations (in state $N$) out of total number of absorptions (either in state $0$ or $N$) over $10^6$ independent realizations of the process. 
The networks in the legend are drawn in the same color as the respective fixation probability.
The inset shows a zoom into the region $0.95 < r <1.05$, where the fixation probabilities are plotted as differences to the well-mixed case. }
\label{fig:PhiPlot}
\end{figure}

The fixation probability on the six different networks of size four is shown in Fig.\ \ref{fig:PhiPlot}. 
It can be seen that, compared to the well-mixed population, the star increases fixation probability for advantageous mutants and decreases it for disadvantageous mutants. 
Thus, the star is called an amplifier of selection.
As the ring is isothermal, i.e. every node has the same number of neighbours, a mutant has the same fixation probability on the ring as in the well-mixed population \citep{lieberman:Nature:2005}.
From Fig.\ \ref{fig:PhiPlot}, we can conclude that the diamond, the shovel, the line and the star are amplifiers of selection.
Thus for size four, all non-isothermal networks are amplifiers, which means that there are no suppressors of selection.
Calling a network an amplifier of selection seems to imply that evolution proceeds faster on this network than on the complete graph.
This arises from the idea that the waiting time for a successful mutation to occur is much longer than the fixation time which the mutation needs to spread through the population. 
In the following, we focus on the fixation time in order to see how it is affected by the removal and addition of links.

\subsubsection{Fixation times}
\label{sec:fixTimes}

With the approach from Section 1\ref{subsec:Grinstead}, we can calculate the expected conditional fixation time for the well-mixed population depending on the fitness $r$ of the mutants,
\begin{equation}
 \tau_{\textrm{mix}} = \frac{11 r^2 + 14r+11}{2( r^2+1)} \ \approx \ 9 - \frac{7}{4} (r-1)^2,
\end{equation}
which is of course identical to the result arising from Eq.\ \eqref{eq:condTime}. 
The ring has the conditional fixation time
\begin{equation}
 \tau_{\textrm{ring}} = \frac{2(3r^2+4r+3)}{r^2+1} \ \approx \ 10 - 2 (r-1)^2 \ .
\label{eq:ring}
\end{equation}
The fixation time $\tau_{\textrm{diamond}}$ can be calculated in the same fashion, but it is a rational function with both numerator and denominator of degree $15$  with up to 13-digit coefficients. 
Therefore, only the Taylor approximation for weak selection up to second order is included here,
\begin{equation} 
  \tau_{\textrm{diamond}}  \ \approx \   10.7\  + \ 0.4(r-1) \ -\  2.5 (r-1)^2 \  .
  \label{eq:fiveLinks}
\end{equation}

The structures shovel, line and star have a substantially higher conditional fixation time than the complete graph, the ring and the diamond. 
For neutral evolution, these are $\tau_{\textrm{shovel}}(1) \approx 16$ , $\tau_{\textrm{line}}(1) \approx 21.3$ and $\tau_{\textrm{star}}(1) \approx 23.2$.
This shows that on every network of size four, neutral fixation is slower than on the complete network. 

\begin{figure}[h!]
 \centering
 \includegraphics[width=0.75\textwidth]{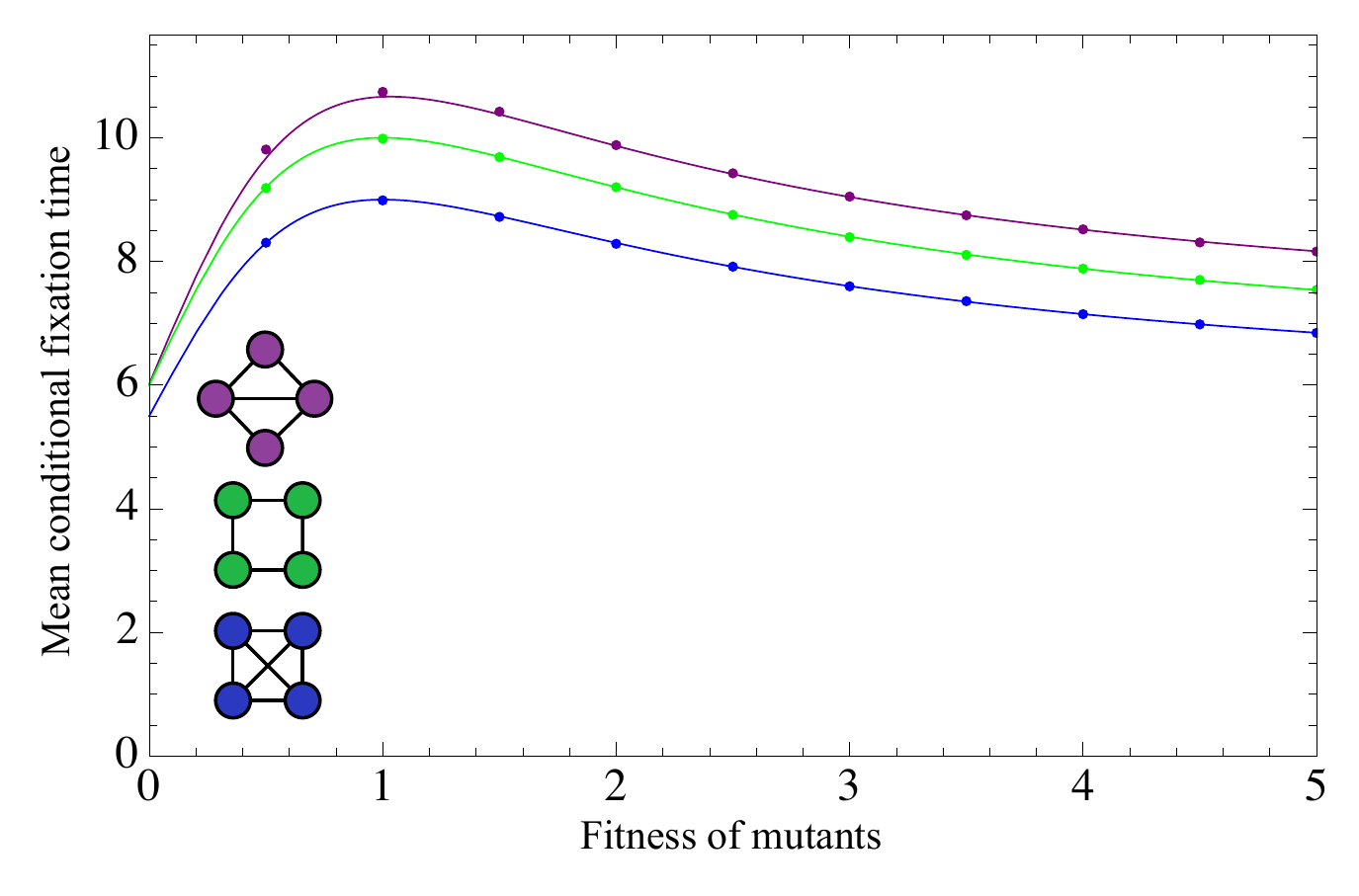}
 \caption{The expected conditional fixation time on three networks of size four. 
 Lines represent the analytical result and each dot is the average over $10^6$ independent realizations.}
 \label{fig:TimeAllGraphsSize4}
\end{figure}

Let us now focus on the complete network, the ring and the diamond to analyse the effect of the removal and addition of one link.
We first compare the analytical results to simulations. 
Fig.\ \ref{fig:TimeAllGraphsSize4} shows the expected conditional fixation time of one advantageous mutant in a population of size four. 
The analytical result is plotted together with the averages over  $10^6$ independent realizations. 
 
Fig.\ \ref{fig:TimeAllGraphsSize4} shows that on the diamond and on the ring, fixation time is higher than on the complete graph. 
On the ring, the increase is approximately one time step, whereas the diamond increases fixation time even more.
Thus, Fig.\ \ref{fig:TimeAllGraphsSize4} reveals a surprising aspect of the fixation times:
intuitively, one may speculate that the removal of a link should always prolong the process of fixation,
because fewer possible paths for the mutant to spread should slow down the process. 
This intuition can be illustrated by road networks, where one would think that more connections speed up overall traffic.
But this is not the true for our case and even for road networks, there are paradoxical situations, where the traffic flow can be increased by closing a road \cite{roughgarden:book:2005}.
From Eqs.\ \eqref{eq:ring} and \eqref{eq:fiveLinks}, we see that although the ring is constructed by dropping one link from the diamond, fixation is faster on the ring than on the diamond. 
This is also seen in Fig.\ \ref{fig:TimeAllGraphsSize4}, where the fixation time is plotted against the mutant's fitness $r$.
This result seems counterintuitive and needs a closer investigation, which can be provided by analysing the sojourn times. 

\subsubsection{Sojourn times}
Intuitively, the more links a network has, the faster the fixation of mutants should happen. 
But this is not true in general, as shown in Fig.\ \ref{fig:TimeAllGraphsSize4}. 
An additional link may not only provide more possibilities for the mutants to place their offspring; 
it may also delay fixation when the mutants replace each other more often.
We look at the sojourn times to understand this process in more detail.

The sojourn times measure how much time is spent in the transient states on average before one of the absorbing states of the system is reached \citep{ewens:book:2004}.
Omitting the cases in which the mutants become extinct, the conditional sojourn times measure the expected time spent in the transient states before mutant fixation.
Summing over the conditional sojourn times before absorption into the all-mutant state yields the total time it takes to go from one to $N$ mutants, the conditional fixation time. 

By comparing the conditional sojourn times in the transient states of different networks, 
we can infer which states cause the delayed fixation.

\begin{figure}[h!]
 \centering
 \includegraphics[width=0.7\textwidth]{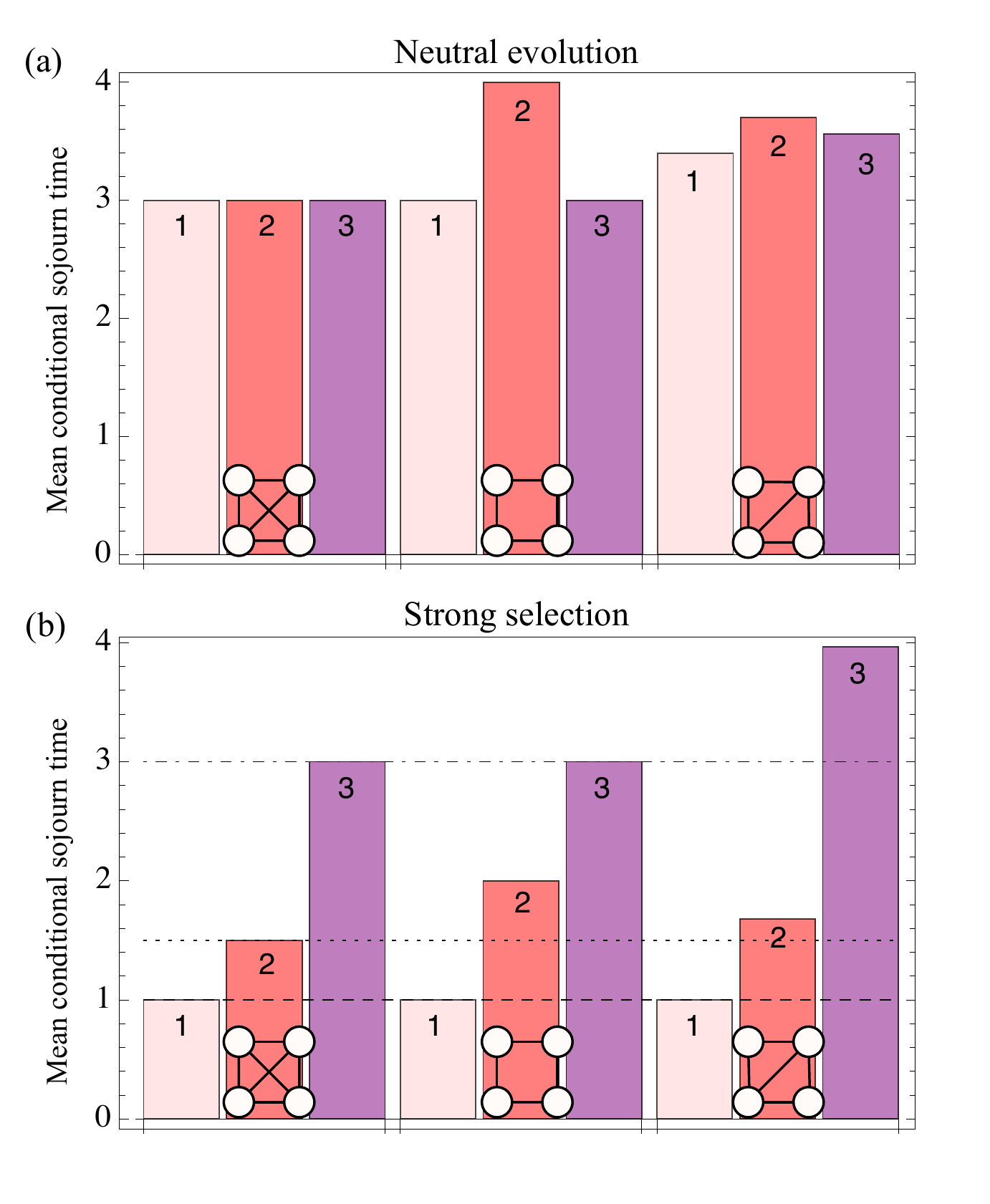}
 \caption{(a): The expected conditional sojourn time for neutral selection $(r=1)$ on the complete graph, the ring and the diamond. 
 (b): The expected conditional sojourn time for strong selection $(r\to \infty)$.
 The numbers on the bars indicate the number of mutants. 
 Note that on the diamond there are several different states for each number of mutants and only the sum of the sojourn times in those states is shown here.}
 \label{fig:SojournBoth}
\end{figure}

In Fig.\ \ref{fig:SojournBoth}, we plot the expected time the system spends in states with one, two and three mutants for the well-mixed population, the ring and the diamond.
For neutral evolution, $r=1$, the well-mixed population and the ring have exactly the same sojourn time in the states with one and three mutants. 
The ring prolongs the sojourn in the state with two mutants for one time step on average. 
The diamond has a shorter sojourn time in the two-mutant states than the ring; however, it increases the sojourn times in all states compared to the complete graph.

In Fig.\ \ref{fig:SojournBoth} (b), we consider the limit of strong selection, $r \to \infty$. 
The process on the ring stays in the states with one and three mutants for the same amount of time as the process on the complete graph. 
So the only difference is the sojourn in the two-mutant state, which causes the conditional fixation time to be higher on the ring than in the well-mixed population.
We saw in Fig.\ \ref{fig:TimeAllGraphsSize4} that a mutant on the diamond has a higher conditional fixation time than on the ring. 
As we can see in Fig.\ \ref{fig:SojournBoth} (b), this increase in fixation time is mainly due to the increased sojourn time in the two different states with three mutants.
The three states with two mutants on the diamond have a lower sojourn time than the two-mutant state on the ring. 
However, the average sojourn in states with three mutants is almost one time step longer.
Interestingly, under strong selection more time is spent in these two states than under neutral selection. This is because the probability to leave states VI and VII decreases with higher intensity of selection. 

So far, we only know that the total sojourn time in all of the three-mutant states is increased on the diamond. 
Let us now address the question as to which of the two different states with three mutants causes the prolonged sojourn time, VI or VII. 
For this purpose, we visualize the sojourn time as a function of the mutant's fitness for all seven transient states.

\begin{figure}[h!]
 \centering
 \includegraphics[width=0.8\textwidth]{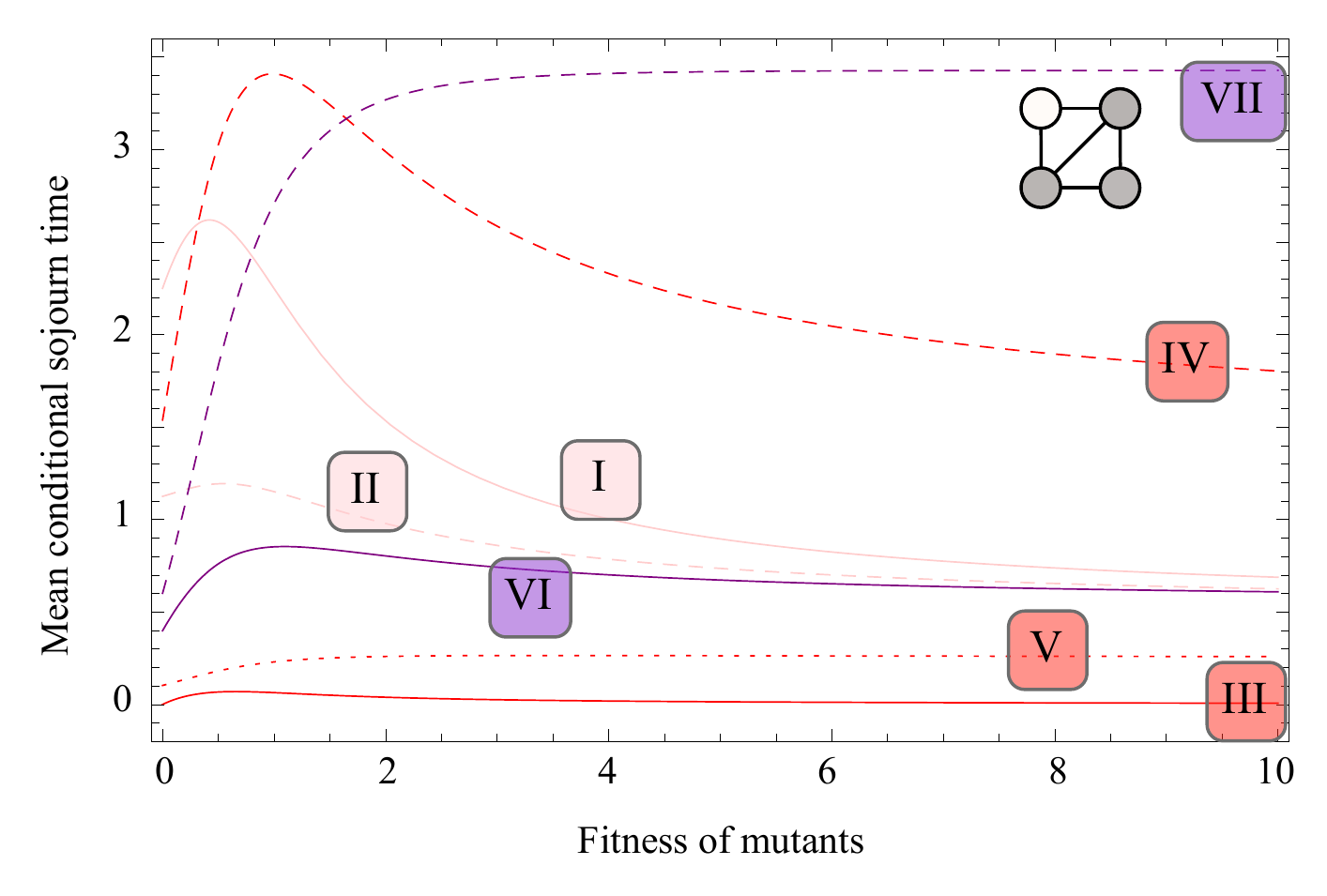}
 \caption{The expected conditional sojourn time in the seven transient states of the diamond. Here, the sojourn time is plotted as a function of the mutant's fitness $r>0$. 
 States I and II have one mutant, states III, IV and V correspond to two mutants and states VI and VII have three mutants (see Fig.\ \ref{graph:StatesFiveLinks})}
 \label{fig:SojournDetailedFiveLinks}
\end{figure}

Looking at the sojourn time in all transient states, see Fig.\ \ref{fig:SojournDetailedFiveLinks}, we see that for neutral evolution and slight fitness difference, the system spends most time in state IV.
This changes at $r \approx 1.65$.
For higher selective advantage, most time is spent in state VII (where the one wild-type individual is situated at a node with two neighbours, see Fig.\ \ref{graph:StatesFiveLinks}.
This can be explained in the following way:
In state VII, the one wild-type individual has lower chances of being replaced, since it is only connected to two of the three mutants.
The three mutants keep replacing each other for a longer time than in state VI, therefore the process stays longer in this state before going to fixation. 

\subsubsection{Initial mutant placement}
\begin{figure}[h!]
 \centering
 \includegraphics[width=0.75\textwidth]{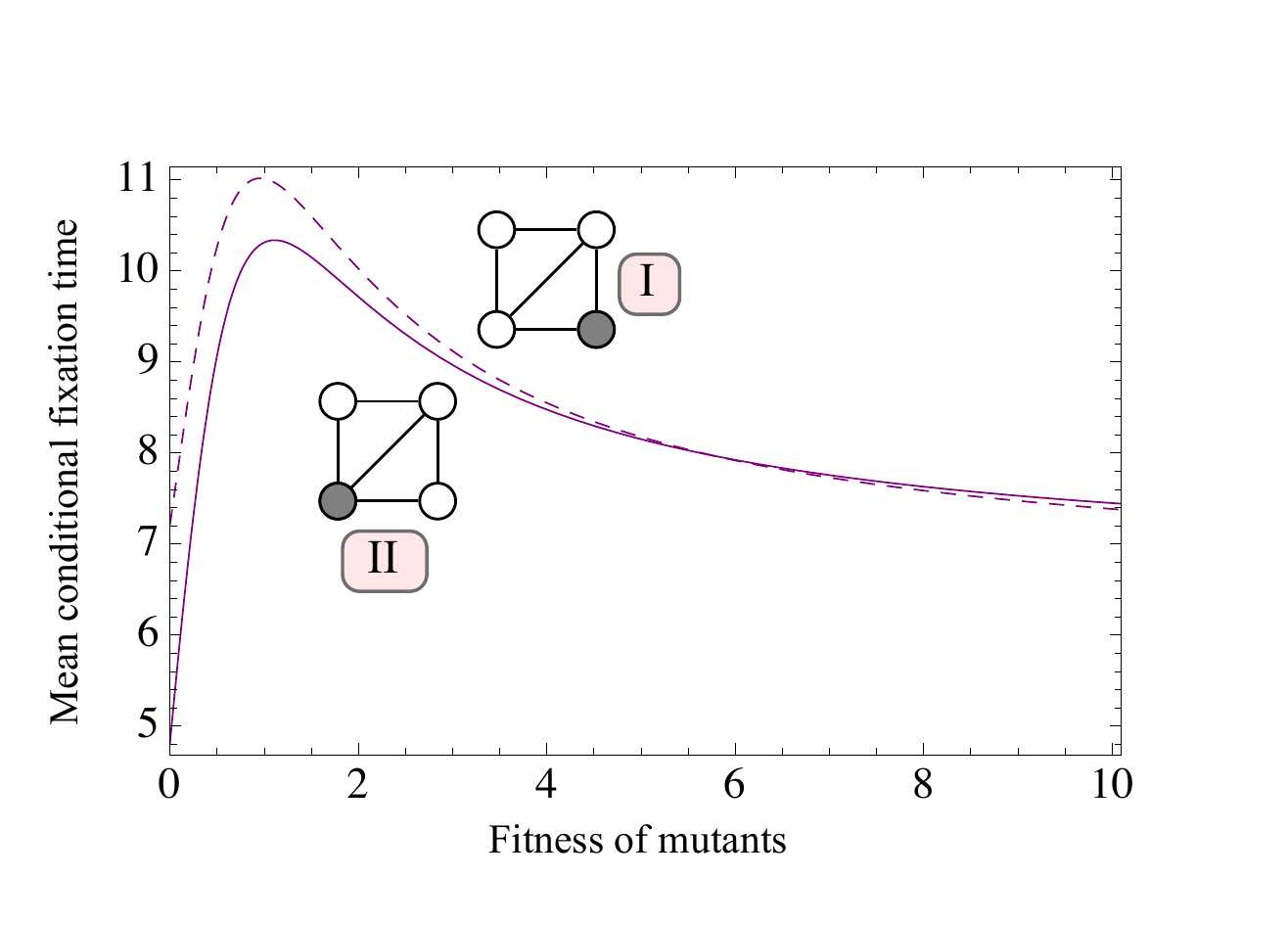}
 \caption{The expected conditional fixation time on the diamond. 
 The first mutant starts either at a node of degree $2$ (dashed line) or at a node with three neighbours (solid line). 
 }
 \label{fig:InitialFiveLinks}
\end{figure}

Instead of randomly choosing a node to place the first mutant on, we now assess the effect of the initial node on the fixation time. 
On the complete graph and the ring, all nodes have the same number of links and therefore all initial conditions are identical.
However, the diamond consists of two nodes with three neighbours and two nodes with two neighbours. 
The conditional fixation time for those two initial conditions is plotted in Fig.\ \ref{fig:InitialFiveLinks}.
The average of the two curves is identical to the conditional fixation time shown in Fig.\ \ref{fig:TimeAllGraphsSize4}, because the probability to place the first mutant at either one of the initial states is $\frac{1}{2}$.

In Fig.\ \ref{fig:InitialFiveLinks}, the initial position of the mutant has a non-trivial impact on the fixation time. 
For small values of $r$, starting from a node with degree $2$ takes longer to fixate than starting from a node with degree $3$. 
This changes at $r \approx 5.8$. 
For larger values of $r$, starting at a node with degree $3$ leads to a slightly higher fixation time than starting at a node with degree $2$.

\subsection{Numerical simulations for larger networks}

\begin{figure}[h!]
 \centering
  \includegraphics[width=0.9\textwidth]{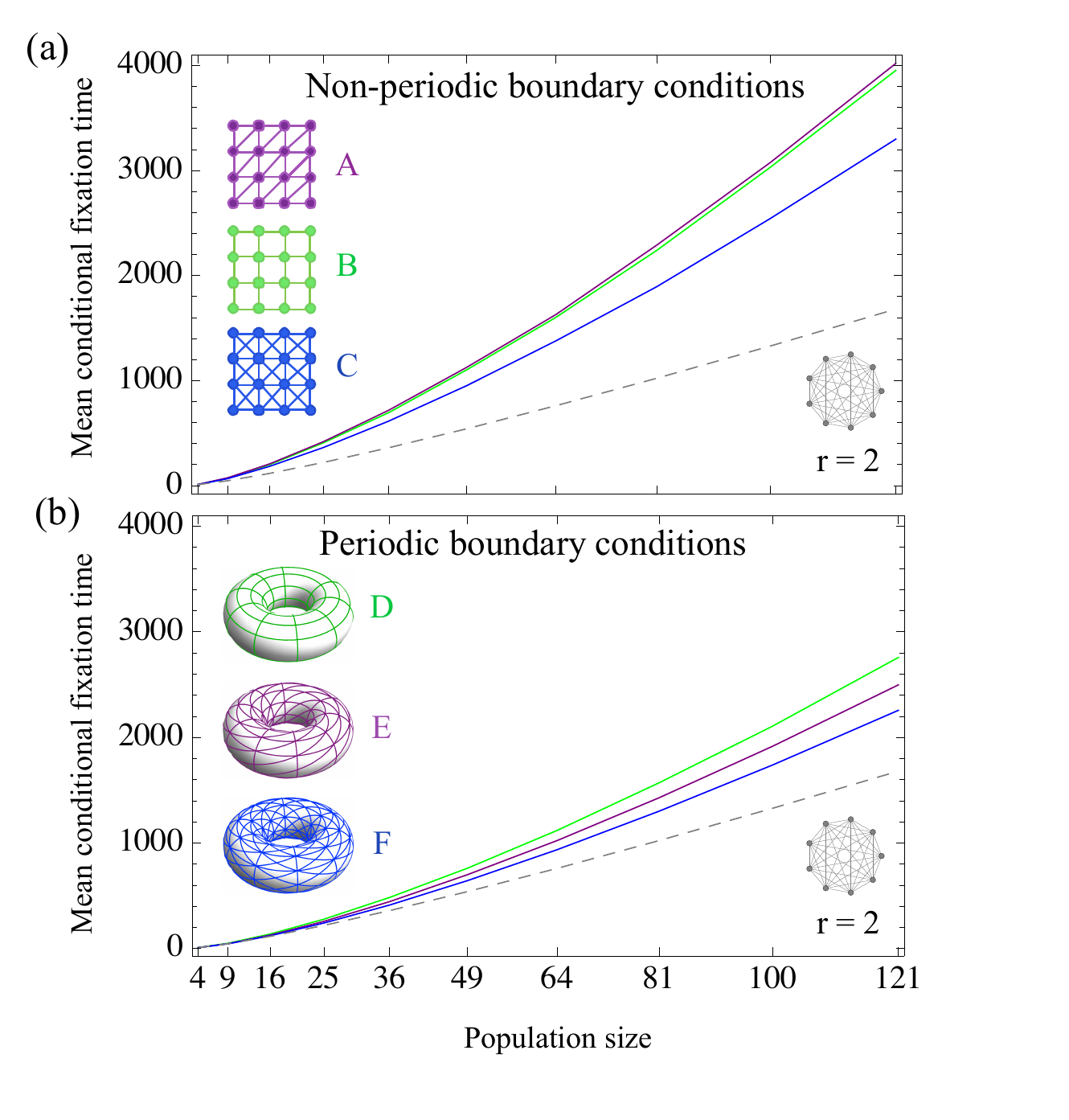}
  \caption{The simulated mean conditional fixation time on three different lattice networks for population sizes $4,9, \dots ,100,121 $ and mutant fitness $r=2$. On both panels, the dashed line depicts the mean conditional fixation time on the complete network of the respective size. 
  Panel (a) shows the fixation time on the lattices without periodic boundary conditions, as they are exemplarily depicted in the legend for size $16$. 
  For the fixation time shown in panel (b), the boundaries of each lattice are connected, such that the structure can be illustrated on the surface of a torus.}
    \label{fig:largerLattices}
 \end{figure}

So far, we have only considered very small networks in order to allow for an analytical consideration of the fixation times. 
Thus, we still have to show that this is not only an effect of these extremely small systems. 
To explore whether the removal of links can also decrease the fixation time, we use our small networks as motifs of a larger network that is constructed from these motifs, see Fig.\ \ref{fig:largerLattices}.
As several links are removed simultaneously, the effect size is not expected to decrease rapidly with the system size. 
For these larger networks, the analytical approach is not feasible and we therefore perform simulations instead. 
We start the birth-death Moran process by putting one mutant on a random node of a network consisting solely of wild-type individuals.
At each time step, one individual is chosen for birth with probability proportional to its fitness.
The offspring then replaces one of its neighbours at random. 
The fixation time of the mutants is averaged over $10^5$ independent realizations.

Fig.\ \ref{fig:largerLattices} shows the mean conditional fixation time on quadratic lattices with and without periodic boundary conditions for mutant fitness $r=2$. 
Fixation occurs much faster with periodic boundary conditions than on the corresponding networks without periodic boundary conditions. 
Note that by connecting the boundaries, we make the lattices isothermal, i.e.\ all nodes have the same number of neighbours. 
Hence, the fixation probability on these networks is the same as on the complete network of the same size.
However, the fixation time is increased, compared with the well-mixed population on a complete network.

Intuitively, the fewer links a graph has, the longer fixation should take.
And as Fig.\ \ref{fig:largerLattices} shows, the lattices with periodic boundary conditions confirm that intuition.
Meaning that starting from network F, by dropping links to obtain network E, the fixation time increases. 
By dropping even more links (network D), the fixation time increases even more.
A similar result was found by Whigham et al. \citep{whigham:GPEM:2008}. 
They simulated the process on different isothermal ring structures of size $N=100$ and showed that the fixation time decreases with increasing node degree. 
But this ordering property by the number of links seems to only hold for isothermal graphs. 

Interestingly, the lattices without periodic boundary conditions do not behave according to that intuition. 
Instead, fixation takes longer on network A than on network B, even though A has more links.
This shows that the results obtained for size $N=4$ are qualitatively still valid for larger lattices without periodic boundary conditions.

 \section{Discussion}
\label{sec:Discussion}
Evolutionary dynamics on networks is a way to consider population structures that
can lead to many non-trivial findings. 
For example, one can construct amplifiers or suppressors of selection which lead to fixation probabilities that deviate from results in a systematic way. 
The isothermal theorem states that for the birth-death Moran process, all regular networks have the same fixation probability, which is a remarkable finding \cite{lieberman:Nature:2005}. 
Also temporal aspects of this dynamics can be highly interesting: a population structure that leads to higher probabilities of fixation can at the same time increase the time of the fixation process itself. 
Based on the results of \cite{frean:PRSB:2013}, we originally set out to show that the fixation time increases on undirected networks if links are removed. 
However, as we have shown here, it is possible to construct counter-examples in which the removal of links decreases the fixation time. 

A similar phenomenon has been reported in transportation systems, where the flow through a system can be increased if connections are removed \cite{gisches:TD:2012}. 
The Braess Paradox describes the situation where adding a seemingly helpful link can have a negative effect on the flow through the network \cite{roughgarden:book:2005}. 

Our analytical approach for small networks allows us to infer how much time the system spends in which states, meaning that the fixation time can be dissected in great detail. 

By analysing the sojourn times, we have shown why our intuition of decreasing fixation time by adding links is not true in general. 
As we have seen, adding links to a network not only increases the possibilities for invasion of the mutants, but also increases the likelihood of mutants replacing each other. 
Therefore, a general statement about the behavior of the fixation time when removing or adding links is not possible.

Furthermore, we have shown that the starting position of the first mutant has a crucial impact on the fixation time.
Depending on the fitness of the mutant, it can be faster if the mutant is placed at a highly connected node or at a node that has only very limited connectivity.
On the diamond, a disadvantageous or slightly advantageous mutant does better in terms of fixation time when starting at the more highly connected nodes. 
However, if the mutant is very advantageous, the more isolated nodes provide a shorter fixation time.

To investigate the counterintuitive result obtained for size $N=4$, the small networks were used to create larger quadratic lattices.
We have shown that this decrease of the fixation time with the removal of links is not just a finite size effect, but can also be found for large population sizes. 

In particular, heterogeneous networks seem to be most relevant for the real world.
They are found among humans and different animal species \cite{wassermann:book:1994,wolf:AB:2007,onnela:PNAS:2007,croft:book:2008,lazer:Science:2009,buettner:PVM:2013,pinter-wollmann:BE:2014}. 
However, to directly transfer our results to such systems seems premature, as it is unclear whether the dynamics in our model
is a good approximation for the processes in these networks. 
Moreover, our analysis reveals that there are still open challenges for a full theoretical understanding of evolutionary dynamics on networks.
Unfortunately, the size of the state space increases rapidly with the system size and requires a tedious construction of transition matrices. 
While analytical approaches are still being developed \cite{adlam:SciRep:2014,maciejewski:JTB:2014,Kaveh:arXiv:2014}, exploring these networks will largely rely on numerical simulations.

\section*{Acknowledgements}
We thank Markus Frean (Wellington University), Paul Rainey (Massey University, Auckland), Benedikt Bauer and Bin Wu for helpful discussions and Bernd Meyer (Monash University, Melbourne) for an important suggestion. 
We thank three anonymous referees for detailed and constructive comments.

\end{document}